\documentclass{ws-procs9x6}

\begin{document}

\title{Lambda and Anti-Lambda Hypernuclei in Relativistic Mean-field Theory}

\author{C. Y. SONG$^*$, J. M. YAO$^*$, H. F. L\"{u}$^{\dag}$ and J. MENG$^{*,\ddag,\S,\P,\|}$}

\address{$^*$ School of Phyics, and State Key Laboratory of Nuclear Physics and Technology,
Peking University, Beijing, 100871, China\\
$^{\dag}$College of Science, China Agriculture University, \\ Beijing 100083, China\\
$^{\ddag}$Institute of Theoretical Physics, Chinese Academy of
Sciences,\\Beijing, 100080, China\\
$^{\S}$Center of Theoretical Nuclear Physics, National Laboratory of
Heavy Ion Accelerator,\\
Lanzhou, 730000, China\\
$^{\P}$Department of Physics, University of Stellenbosch,\\
Stellenbosch, South Africa\\
$^{\|}$E-mail: mengj@pku.edu.cn}

\begin{abstract}
Several aspects about $\Lambda$-hypernuclei in the relativistic mean
field theory, including the effective $\Lambda$-nucleon coupling
strengths based on the successful effective nucleon-nucleon
interaction PK1, hypernuclear magnetic moment and
$\bar\Lambda$-hypernuclei, have been presented. The effect of tensor
coupling in $\Lambda$-hypernuclei and the impurity effect of
$\bar\Lambda$ to nuclear structure have been discussed in detail.

\end{abstract}


\bodymatter

\section{Introduction}\label{aba:sec1}
Since the first discovery of $\Lambda$-hypernuclei by observing
cosmic-rays in emulsion chambers~\cite{PM44348}, lots of efforts
have been devoted to study hypernuclei. Using a variety of
hypernucleus production reactions and coincidence measurement
techniques, data on the
single-$\Lambda$\cite{prsla426,prl2585,prc044302,
Davis86,Hotchi01,Pile91,Nagae00,Hasegawa96,Tamura02,Ukai04,Ajimura01}
and double-$\Lambda$
hypernuclei\cite{np121,prl782,ptp1287,prl132504,prl212502} have been
accumulated. With the additional degree of freedom of strangeness,
hyperons can penetrate into dense nuclear matter inaccessible to
proton and neutron. In astrophysics, hyperons also play a
significant role in the formation and thermal structure evolution of
neutron stars\cite{npa433c,ctp446}. Prospect for production of
neutron halo hypernuclei has been made via $(K^-,\pi^+)$
reaction~\cite{npa211c}, which may be helpful to form neutron halo.

On the theoretical side, non-relativistic few-body model and shell
model as well as the Skyrme-Hartree-Fock theory have been
successfully used to describe single-$\Lambda$ and double-$\Lambda$
hypernuclei\cite{prc041001,Dalitz78,prc3351}. Moreover, the
relativistic mean field (RMF) theory, which was one of the most
successful approaches for ordinary
nuclei\cite{anp1,rpp439,ppnp193,Meng06}, has also been applied to
describe the structure of nuclei with single-$\Lambda$ or
multi-$\Lambda$ and other strange baryons
systems\cite{plb93,npa365,jpg143,plb181,prc2469,npa589,prc322,prc2472,ap35,npa305,prcr1060}.
Particularly, with the relativistic continuum Hartree-Bogoliubov
(RCHB) theory, which has been successfully used to describe the
giant halos in exotic Zr and Ca isotopes\cite{prl460,prcr041302},
the hyperon halo in Carbon hypernuclei and neutron halo in Calcium
hypernuclei have been predicted~\cite{cpl1775,EPJA0305}.


Motivated by the accumulated data of $\Lambda$ binding energy and
spin-orbit splitting, recently the effective $\Lambda$-nucleon
coupling strengths based on the successful effective nucleon-nucleon
interaction PK1 have been proposed with microscopic correction for
the center-of-mass motion. Here the new effective $\Lambda$-nucleon
coupling strengths and calculated results for hypernuclear magnetic
moment and spin symmetry in single $\bar\Lambda$ spectra will be
presented and the effect of tensor coupling in $\Lambda$-hypernuclei
as well as the impurity effect of $\bar\Lambda$ will be discussed in
detail.

\section{Brief introduction of the RMF theory for hypernuclei}

The starting point of the RMF theory is a standard Lagrangian
density ${\cal L}$, in which nucleons are described as Dirac
particles that interact via the exchange of scalar $\sigma$, vector
$\omega$, and isovector-vector $\vec\rho$ mesons as well as
photon~\cite{Meng06}. For hypernuclei system, the Lagrangian density
${\cal L}$ can be written into two parts,
 \begin{eqnarray}
  \label{Eq:Lag}
   {\cal L} = {\cal L}_0 + {\cal L}_Y,
  \end{eqnarray}
where ${\cal L}_0$ is the standard Lagrangian density\cite{Meng06}.
The Lagrangian density ${\cal L}_Y$ for hyperon $Y$
 ($\Lambda$ or $\bar\Lambda$) is given by,
 \begin{eqnarray}
  \label{Eq:Lag_Hyperon}
   {\cal L}_Y
    &=& \bar\psi_Y
      \left(  i\gamma^\mu\partial_\mu - m_Y
               - g_{\sigma Y} \sigma
               - g_{\omega Y} \gamma^\mu\omega_\mu \right)\psi_Y
               + \frac{f_{\omega YY}}{4m_\Lambda}
                 \bar\psi_Y \sigma^{\mu\nu}\Omega_{\mu\nu}\psi_Y,
 \end{eqnarray}
 where $m_Y$ is the mass of hyperon, and $g_{\sigma Y}, g_{\omega Y}$ are the coupling
 strengthes of hyperon and mesons.
 The last term in (\ref{Eq:Lag_Hyperon}) is due to the
 tensor coupling between hyperon and $\omega$ field,
 where the field tensor $\Omega_{\mu\nu}$ for the $\omega$-meson
 is defined as $\Omega_{\mu\nu}=\partial_\mu\omega_\nu-\partial_\nu\omega_\mu$.

%

Restricted to mean field and no-sea approximation, following the
standard procedure\cite{anp1}, one can obtain the equations of
motion for baryons (B) (nucleon (N) and hyperon (Y)) and mesons
respectively, i.e., the Dirac and Klein-Gordon equations.

The Dirac equation for the hyperon is,

 \begin{eqnarray}
 \left[\mathbf{\alpha}\cdot \mathbf{p}+(m_Y + S_Y)
 + \gamma_\mu V^\mu_Y
 -  \frac{f_{\omega YY}}{2m_Y} \sigma_{\mu\nu} \partial^\mu \omega^\nu \right]
 \psi_{Y i}(\bm{r})=\epsilon_i\psi_{Y i}(\bm{r}),
 \end{eqnarray}
 with the vector potential $\displaystyle V^\mu_Y  = g_{\omega Y}
\omega^\mu$ and scalar potential $S_\Lambda  =  g_{\sigma Y}
\sigma$. For the $\omega$ meson, the corresponding Klein-Gordon
equation reads,
  \begin{eqnarray}
(-\nabla^2 + m^2_\omega)\omega_\mu
 = \sum_{B}g_{\omega B} j^B_\mu
   - c_3 \omega_\nu\omega^\nu\omega_\mu
   - \frac{f_{\omega YY}}{2m_Y}j^{T,Y}_\mu,
  \label{Eq:motion}
 \end{eqnarray}
where the baryon current $j^B_\mu$ and tensor current $j^{T,Y}_\mu$
have been respectively defined as,
 \begin{eqnarray}
 j^B_\mu            &=& \sum_{i}\bar\psi_{Bi}\gamma_\mu\psi_{Bi},\\
 j^{T,Y}_\mu&=&\sum_{i}\partial^\nu(\bar\psi_{Y i}  \sigma_{\mu\nu}\psi_{Y i}).
 \end{eqnarray}
More details can be found in Ref.~\cite{Meng06}.


\section{New hyperon-nucleon parametrization}

The RMF theory has made great success in the description of ordinary
nuclei with an universal effective nucleon-nucleon interaction
\cite{npa557,prc540,prc034319} determined by fitting the nuclear
observables such as binding energy, charge radii, etc. As PK1 is one
of the most successful effective nucleon-nucleon interaction, it is
natural to extend PK1 for the description of hypernuclei.

For $\Lambda$-nucleon effective interaction, there are four
additional parameters $m_\Lambda$, $g_{\sigma\Lambda}$,
$g_{\omega\Lambda}$ and $f_{\omega\Lambda\Lambda}$. The mass of
$\Lambda$ is usually fixed to the experimental value
$M_{\Lambda}=1115.6$ MeV. As suggested in Ref.\cite{Cohen91}, the
tensor $\omega$-$\Lambda$ is adopted as $R_{\omega\Lambda\Lambda} =
f_{\omega\Lambda\Lambda}/g_{\omega\Lambda} = 1$. The other two
parameters are usually determined by fitting the single-$\Lambda$
binding energy and /or the spin-orbit splitting.
%

In Ref.~\cite{Lv09I}, a new effective hyperon-nucleon interaction
Y1, based on the effective nucleon-nucleon interaction PK1, has been
developed and labeled as PK1-Y1 by fitting the single-$\Lambda$
binding energies of hypernuclei $^{12-14}_{~~~~\Lambda}$C,
$^{15}_{\Lambda}$N, $^{16}_\Lambda$O, $^{28}_\Lambda$Si,
$^{32}_\Lambda$S, $^{40}_\Lambda$Ca, $^{51}_\Lambda$V,
$^{89}_\Lambda$Y, $^{139}_\Lambda$La and $^{208}_\Lambda$Pb as well
as the spin-orbit splittings in ${}_\Lambda^{9}$Be and
${}_\Lambda^{13}$C.
 The ratio of the meson-$\Lambda$ coupling strengthes to meson-nucleon
 ones, i.e., $R_\sigma = g_{\sigma \Lambda}/g_{\sigma N} $
 and $R_\omega = g_{\omega \Lambda}/g_{\omega N}$, thus obtained are
 $R_\sigma =
0.580$ and $R_\omega = 0.620$. The parameters of effective
$\Lambda$-nucleon interactions PK1-Y1 is shown in
Table~\ref{table:1} in comparison with TM1-B\cite{npa557}, NLSH-A,
and NLSH-B\cite{plb377}. The root-mean-squared (rms) deviation
$\Delta$ and the relative ones $\chi$ for single-$\Lambda$ binding
energies and the $\Lambda$ spin-orbit splitting of $p$ state in
${}_\Lambda^{9}$Be and ${}_\Lambda^{13}$C are also presented, where
$\Delta\equiv \sqrt{\dfrac{1}{N} \sum_{i=1}^N ({\cal O}_i^{\rm
exp.}-{\cal O}_i^{\rm theo.})^2}$ and $\chi\equiv \sqrt{\dfrac{1}{N}
\sum_{i=1}^N \dfrac{({\cal O}_i^{\rm exp.} - {\cal O}_i^{\rm
theo.})^2}{({\cal O}_i^{\rm exp.})^2}}$.

\begin{table}[h!]
\tbl{The parameters of effective $\Lambda$-nucleon interactions
PK1-Y1, TM1-B, NLSH-A, and NLSH-B as well as the corresponding
root-mean-squared (rms) deviation $\Delta$ and the relative ones
$\chi$ for single-$\Lambda$ binding energies of candidate
hypernuclei ($\Delta_b$, $\chi_b$) and the $\Lambda$ spin-orbit
splitting of $p$ state in ${}_\Lambda^{9}$Be and ${}_\Lambda^{13}$C
($\Delta_p$, $\chi_p$).}
 {\begin{tabular}{@{}cccccc@{}}
  \toprule
  Sets & PK1-Y1          &  TM1-B          &  NLSH-A       & NLSH-B \\
   \colrule
 $R_\sigma,R_\omega$         & 0.580,0.620    & 0.468,0.485   &  0.621,0.667    &  0.490,0.512  \\
 $R_{\omega\Lambda\Lambda}$  & 1.0            & 1.21          & 1.0             &  0.616 \\
 \hline
   $\Delta_b$         & 0.851          & 1.229      & 1.200  & 0.906    \\
  $\chi_b(10^{-2}) $  & 5.419          & 6.164      & 7.539  & 5.470    \\
   $\Delta_p$         & 0.058          & 0.103      & 0.100  &  0.257    \\
   $\chi_p$           & 0.391          & 0.915      & 0.682  &   1.698    \\
 \botrule
 \end{tabular}}
 \label{table:1}
 \end{table}

 \begin{figure}[h!]
 \centering
 \psfig{file=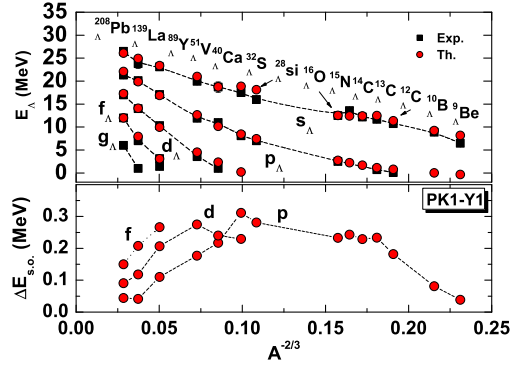,width=3.0in} \caption{Single-$\Lambda$
 binding energies (upper panel) and spin-orbit splitting sizes (lower panel)
 for $\Lambda$ states in RMF calculations with newly-adjusted PK1-Y1 effective interaction.
 For comparison, the experimental data\cite{Davis86,Hotchi01,Pile91,Nagae00,Hasegawa96,Tamura02,Ukai04,Ajimura01} are given as well.}
 \label{aba:fig1}
 \end{figure}

 Figure~\ref{aba:fig1} shows the single-$\Lambda$ binding energies
 and spin-orbit splitting sizes for $\Lambda$ states with different
 orbital angular momentum obtained from RMF
 calculations with PK1-Y1 effective interaction.
 In the upper panel, good agreement for single-$\Lambda$ binding energies
 has been achieved by the theoretical calculations.
 For the $\Lambda$ spin-orbit splitting, different from that for
 nucleon, the magnitude of around
 several hundreds keV has been found in the calculation.
 The splitting in medium-mass
 region are relatively larger than those in
 light- and heavy- mass regions.

\section{Magnetic moments of  $\Lambda$-hypernuclei}

With fast development of experimental techniques, the interest in
hypernuclear magnetic moments is evoked. The effects of core
polarization and tensor coupling on the magnetic moments in
$^{13}_\Lambda$C, $^{17}_\Lambda$O, and $^{41}_\Lambda$Ca
hypernuclei are studied in the Dirac equation with scalar, vector
and tensor potentials~\cite{Yao08}. It is shown that the inclusion
of a $\Lambda$ tensor coupling will modify the current vertex and
suppress the effect of core polarization on the magnetic moments.
However, as the hyperon wave functions are not sensitive to the
$\Lambda$ tensor potential, the magnetic moments with or without
$\Lambda$ tensor potential are almost the same. The deviations of
magnetic moments for $\Lambda$ in $p$ states from the Schmidt values
are found to increase with the nuclear mass number.

However, this study is based on the perturbation theory for the
symmetric nuclear matter. A self-consistent calculation in finite
hypernuclei is required, in which both the nucleons and hyperon are
treated on the same footing. A self-consistent time-odd triaxial RMF
approach~\cite{Yao06} include the hyperon and the tensor coupling is
developed and applied to study the magnetic moments in
hypernuclei~\cite{Lv09II}. The magnetic moments of
$^{16-}$$^{18}_\Lambda$O by time-odd triaxial RMF approach with PK1
and PK1-Y1 are shown in Table~\ref{table:3}. It is found that the
core polarization effect of valence $\Lambda$ is very important
although it is smaller than that of the valence neutron.
Furthermore, the core polarized Dirac magnetic moment might be
reduced by the tensor coupling of the valence $\Lambda$.

 \begin{table}[h!]
 \tbl{The magnetic moments of oxygen hypernuclei in units of nucleon magneton (n.m.),
  by time-odd triaxial RMF approach with PK1 and PK1-Y1.
  The total magnetic moment $\mu_{\rm tot.}$ is given by
  the sum of  $\mu_D$, the anomalous magnetic moment of the nuclear
  core $\mu^{\rm n+p}_a$ and hyperon magnetic moment
  $\mu_a^\Lambda$. While the Schmidt magnetic moment is represented
  by $\mu_{\rm Sch.}$.}
   { \begin{tabular}{@{}cccccc@{}}\toprule
    Sys. & $\mu_D$ & $\mu^{\rm n+p}_a$ & $\mu_a^\Lambda$ &  $\mu_{\rm tot.}$  & $\mu_{\rm Sch.}$  \\
    \colrule
    $^{15}$O + free~$\Lambda$& -0.113   & 0.677 & -0.613 & -0.049 &  0.025   \\
    $^{16}_\Lambda$O(Y1)     & -0.132              & 0.681              & -0.610 & -0.060 &  0.025   \\
    \colrule
    $^{16}$O + free~$\Lambda$&  0.    & 0. & -0.613 & -0.613 &  -0.613  \\
    $^{17}_\Lambda$O(Y1)     & -0.005 & 0. & -0.610 & -0.614 &  -0.613  \\
  \colrule
    $^{17}$O + free~$\Lambda$& -0.134  & -1.863  & -0.613 & -2.610 & -2.526   \\
    $^{18}_\Lambda$O(Y1)     & -0.146              & -1.862              & -0.610 & -2.618 & -2.526   \\
    \botrule
 \end{tabular}}
 \label{table:3}
\end{table}

\section{Nucleus with anti-Lambda}

In Ref.\cite{PRL91262501}, the anti-nucleon spectrum has been
studied for ordinary nuclei with the RMF theory and the spin
symmetry is found for the single anti-nucleon spectra, i.e., the
spin partner states are nearly degenerate and the dominant
components of the wave functions are almost the same. It is
therefore worthwhile to examine the spin symmetry in single
$\bar{\Lambda}$ spectra.

By taking $^{16}\mathrm{O}$ system as the representative case, the
single $\bar{\Lambda}$ spectra and the $\bar{\Lambda}$ wave
functions were studied in Ref.\cite{Song09}. In the Dirac equation
of $\bar{\Lambda}$, the scalar and vector potentials of
$\bar{\Lambda}$ are written respectively as $S_{\bar{\Lambda}}(r)=
g_{\sigma \bar{\Lambda}}\sigma$ and $V_{\bar{\Lambda}}(r)=g_{\omega
\bar{\Lambda}}\omega_{0}$. The charge conjugation leaves the scalar
potential invariant, $S_{\bar{\Lambda}}(r)=S_{\Lambda}(r)$, and
changes the sign of the vector potential,
$V_{\bar{\Lambda}}(r)=-V_{\Lambda}(r)$.

\begin{figure}
\centering
\psfig{file=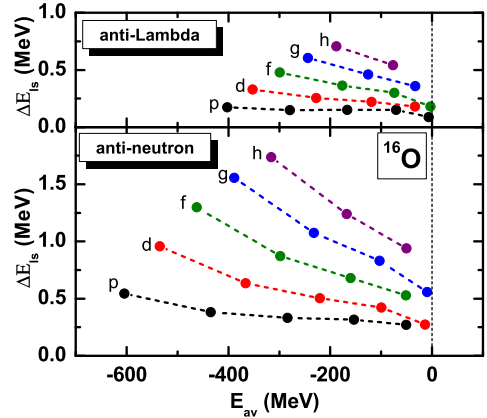,width=2.5in} \caption{Spin-orbit
splitting $\epsilon_{A}(nl_{l-1/2})-\epsilon_{A}(nl_{l+1/2})$ in the
spectra of anti-Lambda and anti-neutron in $^{16}\mathrm{O}$ versus
the average energy of a pair of spin doublets. The vertical dashed
line shows the continuum limit.} \label{aba:fig2}
\end{figure}

In Figure~\ref{aba:fig2} are shown the spin-orbit splittings
$\epsilon_{A}(nl_{l-1/2})-\epsilon_{A}(nl_{l+1/2})$ of anti-Lambda
and anti-neutron as functions of the average energy for spin
partners in $^{16}\mathrm{O}$. The values of splitting for
anti-Lambda ($0.1\sim0.8$ MeV) are smaller than those of
anti-neutron ($0.2\sim1.9$ MeV), which implies that the spin
symmetry in anti-Lambda spectra is even better conserved than that
in anti-neutron spectra. It is also found that the dominant
components of $\bar{\Lambda}$ Dirac spinors are almost identical for
spin partner states.

The self-consistent effects caused by the $\bar{\Lambda}$ had not
been taken into account in the above calculations. For a real
$\bar{\Lambda}$-$^{16}\mathrm{O}$ system, the self-consistent mean
fields including the scalar and vector ones will be modified by the
$\bar{\Lambda}$. Further investigation on this issue is in progress.

 \section{Summary}

Several aspects about hypernuclei investigated with the relativistic
mean field theory, including the effective $\Lambda$-nucleon
coupling strengths based on the effective nucleon-nucleon
interaction PK1, hypernuclear magnetic moment and spin symmetry in
$\bar\Lambda$-hypernuclei have been presented. With the
newly-adjusted  PK1-Y1 effective interaction, the
 single-$\Lambda$ binding energies of hypernuclei from light
 to heavy mass regions have been well reproduced. The effects
 of tensor coupling and core polarization
 from valence $\Lambda$ in $\Lambda$-hypernuclei have been
 found to be of importance in the description
 of hypernuclear magnetic moments.
 The spin symmetry in $\bar{\Lambda}$ spectra have been
 found to be even better developed than that in anti-neutron spectra.
 The investigation for spin symmetry in single-$\bar \Lambda$
 spectra in $\bar{\Lambda}$-hypernuclei is in progress.

 \section*{Acknowledgments}

This work is partly supported by the Major State Basic Research
Development Program (2007CB815000) and the National Natural Science
Foundation of China 10775004.

\end{document}